\documentclass[review,number,5p,sort&compress]{elsarticle}
\usepackage[utf8]{inputenc}
\usepackage{amsmath,amssymb} 
\usepackage{graphicx}
\usepackage{CJKutf8}  
\usepackage{hyperref}
\usepackage[capitalise]{cleveref}   

\begin{document}

\begin{frontmatter}
\title{Upgrade of the MARI spectrometer at ISIS}
\author[label1]{M. D. Le}
\author[label1,label2]{T. Guidi}
\author[label1]{R. I. Bewley}
\author[label1]{J. R. Stewart}
\author[label1]{E. M. Schooneveld}
\author[label1]{D. Raspino}
\author[label1]{D. E. Pooley}
\author[label1]{J. Boxall}
\author[label1]{K. F. Gascoyne}
\author[label1]{N. J. Rhodes}
\author[label1]{S. R. Moorby}
\author[label1]{D. J. Templeman}
\author[label1]{L. C. Afford}
\author[label1]{S. P. Waller}
\author[label1]{D. Zacek}
\author[label1]{R. C. R. Shaw}
\address[label1]{ISIS Neutron and Muon Source, STFC Rutherford Appleton Laboratory, Harwell Campus, Didcot OX11 0QX, United Kingdom}
\address[label2]{Physics Division, School of Science and Technology, University of Camerino, 62032 Camerino, Italy}
\date{November 2021}

\begin{abstract}
The MARI direct geometry time-of-flight neutron spectrometer at ISIS has been upgraded with an $m=3$ supermirror guide and new detector electronics. This has resulted in a flux gain of $\approx$6$\times$ at $\lambda=1.8$~\AA, and improvements on discriminating electrical noise, allowing MARI to continue to deliver a high quality science program well into its fourth decade of life.
\end{abstract}

\end{frontmatter}

\section{Introduction}

MARI is a time-of-flight direct-geometry spectrometer located in the first target station at the ISIS Neutron and Muon Spallation Source, Rutherford Appleton Laboratory, UK.  It was amongst the first instruments designed and built at the early days of ISIS~\cite{taylor90}, and was funded as part of the UK-Japan collaboration initiated by the late Professor Yoshikazu Ishikawa. The instrument was named MARI (\begin{CJK*}{UTF8}{bsmi}真理\end{CJK*}) in honour of Prof. Ishikawa’s daughter, and means “Truth” in Japanese. It was completed and started user experiments at the end of 1989. Its original design complemented the highly successful HET (High Energy Transfer) spectrometer by increasing the detector coverage and having a peak flux at lower energies (at $\approx$25~meV) provided by the 110~K liquid methane moderator. The 4~m secondary flight path also ensures a good energy resolution, which can be as fine as 1\% of the incident energy, and it has a unique vertical scattering geometry giving it an extremely low background count rate of approximately 6 counts per hour per metre of detector.

The 1~m wide detector bank with 10-atm 1$''$ diameter non-position sensitive $^{3}$He detector tubes covering a wide scattering angle range ($\approx 5-135^{\circ}$) has allowed many different types of scientific problems to be addressed by the instrument on polycrystalline and liquid samples in its first two decades of operation. 
While initially focused on measurements of phonon density of states (PDOS) in crystalline solids, liquids and amorphous materials, over time the science program has shifted towards the study of magnetic excitations, to the point that around two-thirds of the beam time is currently devoted to magnetic materials. 

Despite this past success, by modern standards MARI's incident flux was too low to be competitive, as it possessed no neutron guide. It was thus recently upgraded with an $m$=3 straight converging neutron supermirror guide which has increased its flux by a factor of 6 at 25~meV (1.8~\AA), where the peak flux is. This gain is larger at lower energies.
In addition to the new guide, MARI has also had a new double-disk chopper installed to allow greater flexibility in the use of replication rate multiplication (RRM), and the detector and data acquisition electronics were also upgraded, allowing better signal processing and improved event-mode data collection.

\begin{figure*}[hbt!]
  \begin{center}
    \includegraphics[width=0.85\textwidth,viewport=0 0 702 606]{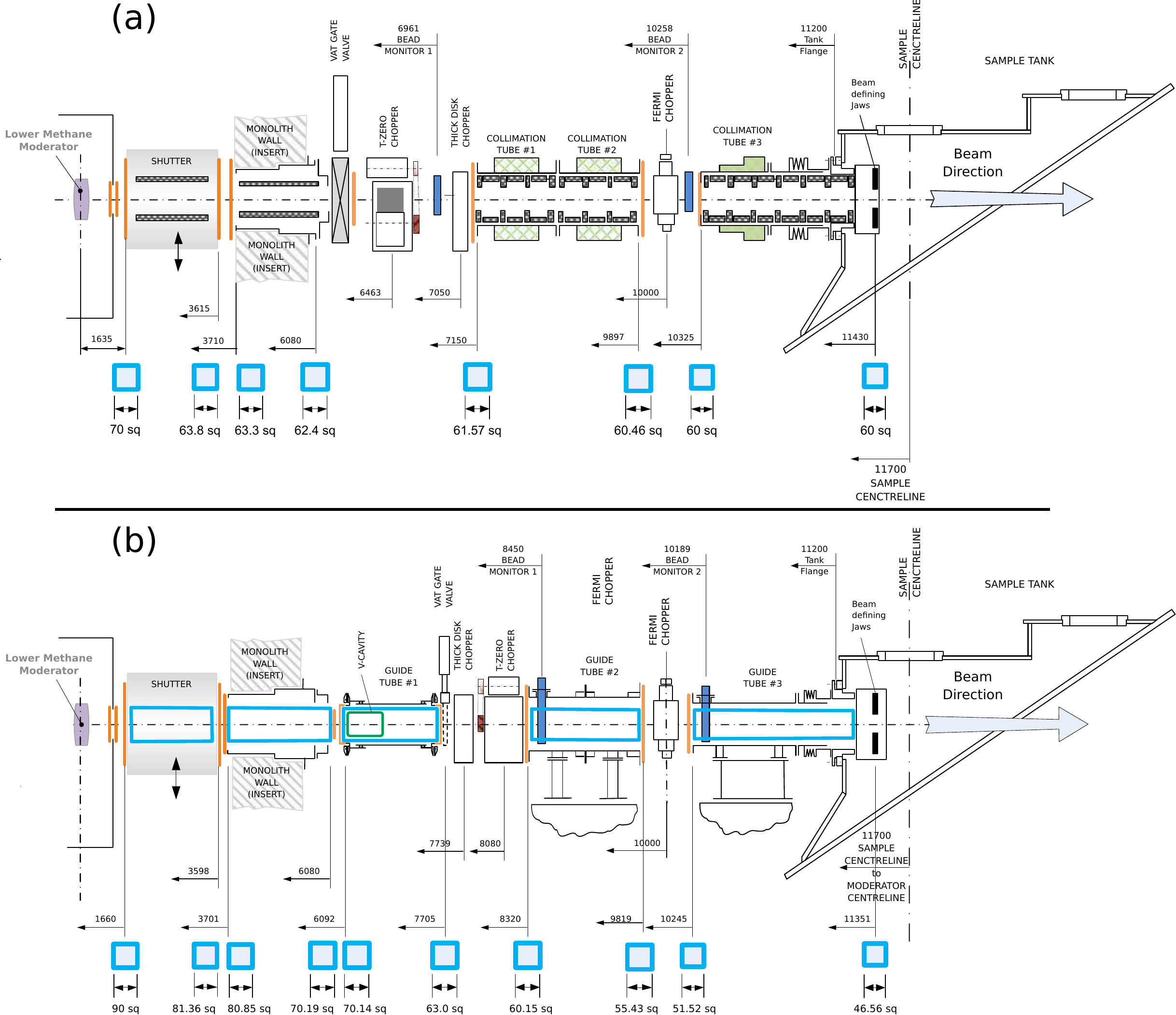}
    \caption{Schematic of (a) the old and (b) the new designs of the MARI spectrometer.
    Blue squares below each schematic indicate the cross-section size of the beam at that point.
    All measurements are in millimetres and distances are specified with respects to the moderator centre unless otherwise stated.}
    \label{fg:overview2}
  \end{center}
\end{figure*}

With several years of user operation since the upgrade, the improved performance of MARI has enabled experiments to be carried out faster and for previously unfeasible experiments to be completed, in a variety of science areas from thermoelectric and battery materials to superconductivity and low dimensional magnetism.

We first describe the original MARI (\cref{sec-oldmari}), then detail the guide (\cref{sec-guide}), choppers (\cref{sec-choppers}) and detector electronics (\cref{sec-det}) upgrades. Finally, we evaluate the performance of the upgraded instrument (\cref{sec-performance}) and showcase recent scientific examples (\cref{sec-science}).

\section{MARI original design} \label{sec-oldmari}

The pre-upgrade layout of the MARI primary spectrometer is shown schematically in~\cref{fg:overview2}(a), and consists of evacuated square cross-sectioned beam tubes lined with neutron-absorbing boron carbide (B$_4$C) together with three choppers: A T0 chopper made from Nimonic alloy at 6.4~m from the moderator, to stop fast neutrons; A thick disk chopper with a single slot at 7.0~m to transmit a single neutron pulse for each proton pulse and hence reduce background and a monochromating fast Fermi chopper at 10~m. The disk chopper was often only used with the gadolinium Fermi chopper package as this could let through ``$\pi$'' pulses close enough in incident energy to the desired neutron pulse to overlap with it, especially when the Fermi was run at high frequencies. The curved slit packages of the other chopper packages (which had thicker boron slats) meant that only ``$2\pi$'' pulses were transmitted and pulse-overlap was not such a large problem. Hence often the disk was stopped and left open for these measurements.

There is a beam monitor (M1) before the disk chopper which can be used to normalise the data, and another monitor after the Fermi chopper (M2), which together with a final monitor in the ``get-lost'' tube before the beam stop (M3, not shown) is used to determine the energy of the neutrons incident on the sample.

Finally, the sample tank is shown which may be pumped down to a cryogenic vacuum ($\approx 10^-7$~mBar), whilst the detectors (not shown) form a circular arc \textit{below} the sample covering scattering angles from -15$^{\circ}$ to +135$^{\circ}$, at a sample-detector distance of 4~m.

\section{Guide upgrade} \label{sec-guide}

The new $m=3$ converging guide is installed in 5 sections, shown schematically as blue rectangles in \cref{fg:overview2}(b).
The first section is embedded in a new evacuated shutter, and the second section is mounted in an insert tube in the target station monolith shielding.
Sections 3-5 are outside the monolith and are separated by two pits, the first containing the disk and Nimonic (T0) choppers and the second with the fast Fermi chopper.
These guide sections have also been fitted with inclinometers to monitor, via a control box and software, the guide alignment over time. One packaged unit is screwed to the outside of each guide's vacuum housing plus an internal unpackaged unit on each internal glass section.
The new Fermi chopper is mounted on a lift and can be moved out of the beam for white beam measurements without manually removing the chopper as was the case before the upgrade.

Whilst the Fermi chopper position remains the same (10~m from the moderator), the disk and Nimonic choppers were moved closer to the sample so as to reduce the possibility of radiation damage. In previous operations, it was found that the motor encoders on these choppers needed to be replaced every 1-2~years.
In addition, due to space constraints and in order to minimise air gaps between components, the new double disk chopper is placed \textit{before} the Nimonic chopper, and the first monitor (M1) is placed in Guide 4 after the disk and Nimonic. As MARI now often runs with the Gd chopper and so must use the disk, this monitor no longer records the white beam incident spectrum. This monitor and the monitor (M2) after the Fermi chopper are placed in vacuum inside guides 4 and 5 respectively.

\begin{figure}
  \begin{center}
    \includegraphics[width=0.99\columnwidth, viewport=18 5 418 315]{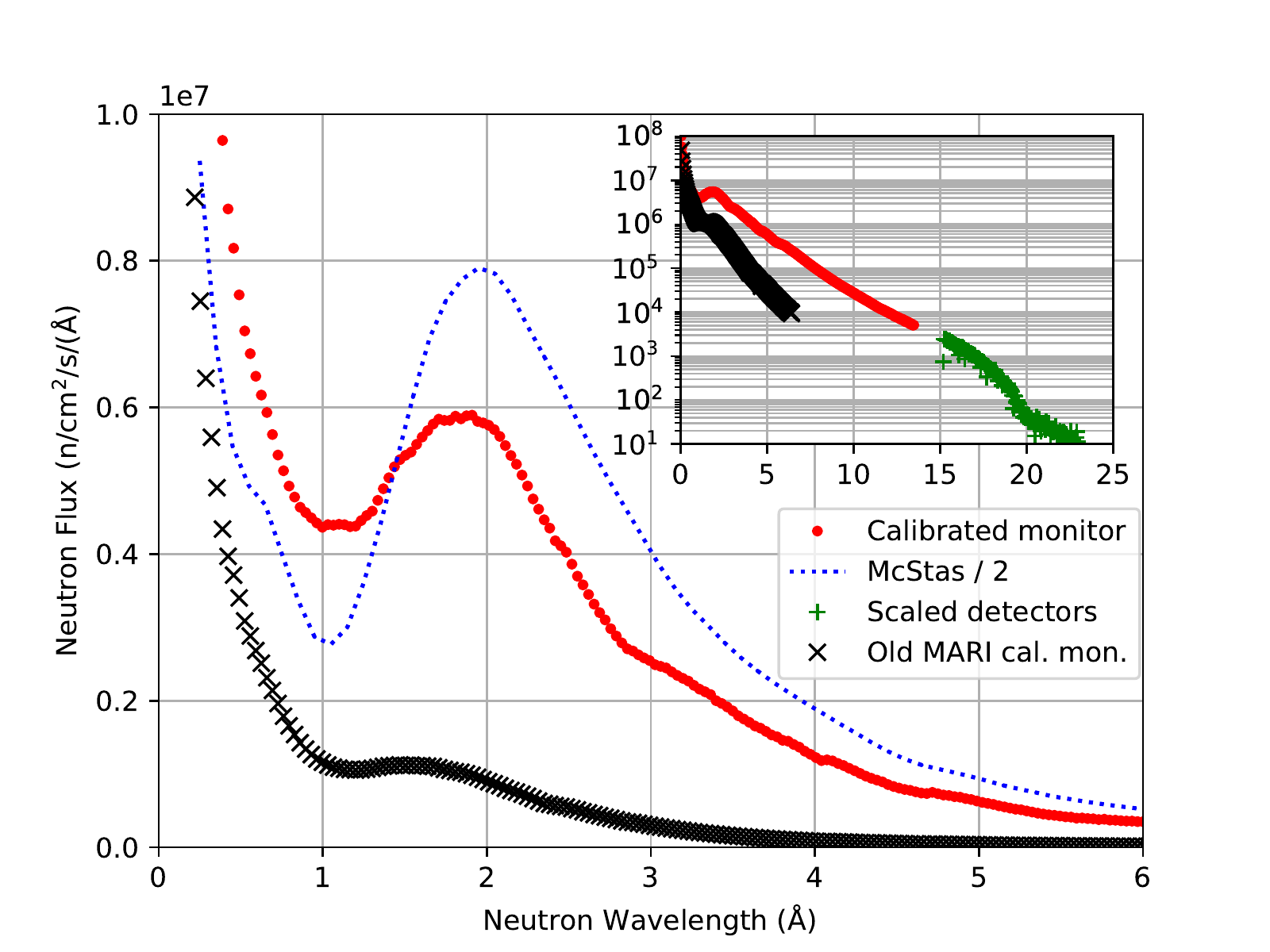}
    \caption{Measured (circles) and calculated (dotted line) white beam flux at the sample position of upgraded MARI normalised to a nominal proton beam current of 175~$\mu$A. The inset shows an expanded spectra with supplementary measurements up to 23~\AA~using the low angle detectors (crosses) as described in the text to illustrate the attenuation cut-off around 19~\AA~due to the V-cavity frame overlap mirror.}
    \label{fg:absflux}
  \end{center}
\end{figure}

The initial opening and view of the methane moderator is $90 \times 90$~mm, and this converges down to a view of $45\times 45$~mm at the sample position, compared with the pre-upgrade cross-sections of $70\times 70$~mm at the moderator end and $60\times 60$~mm at the sample end. The new larger view of the moderator gives a 1.65$\times$ gain at all energies, whereas the gain from the guide itself is neutron wavelength dependent, approaching unity at short wavelengths (high energies).
Section 3 also has a single 200~mm long V-cavity frame-overlap mirror (FOM), consisting of two $m=5$ supermirrors on a Si substrate at a taper of 9$^{\circ}$, designed to scatter neutrons with wavelength longer than $\approx $20~\AA~out of the beam.
The inset to \cref{fg:absflux} shows the long wavelength flux (green crosses), and the reduction due to the FOM around 20~\AA. These measurements were made with the disk chopper running at 10~Hz (without the Fermi chopper) and uses the low angle (3.5$^{\circ}$) detectors, as these neutrons are very divergent ($\approx 5^{\circ}$) so they impinge directly on the low angle detectors. This fact, however, makes the low angle detectors unusable in normal operation, where we find that we need to mask detectors below $\approx 6^{\circ}$.

\begin{figure}
  \begin{center}
    \includegraphics[width=0.99\columnwidth, viewport=24 5 418 320]{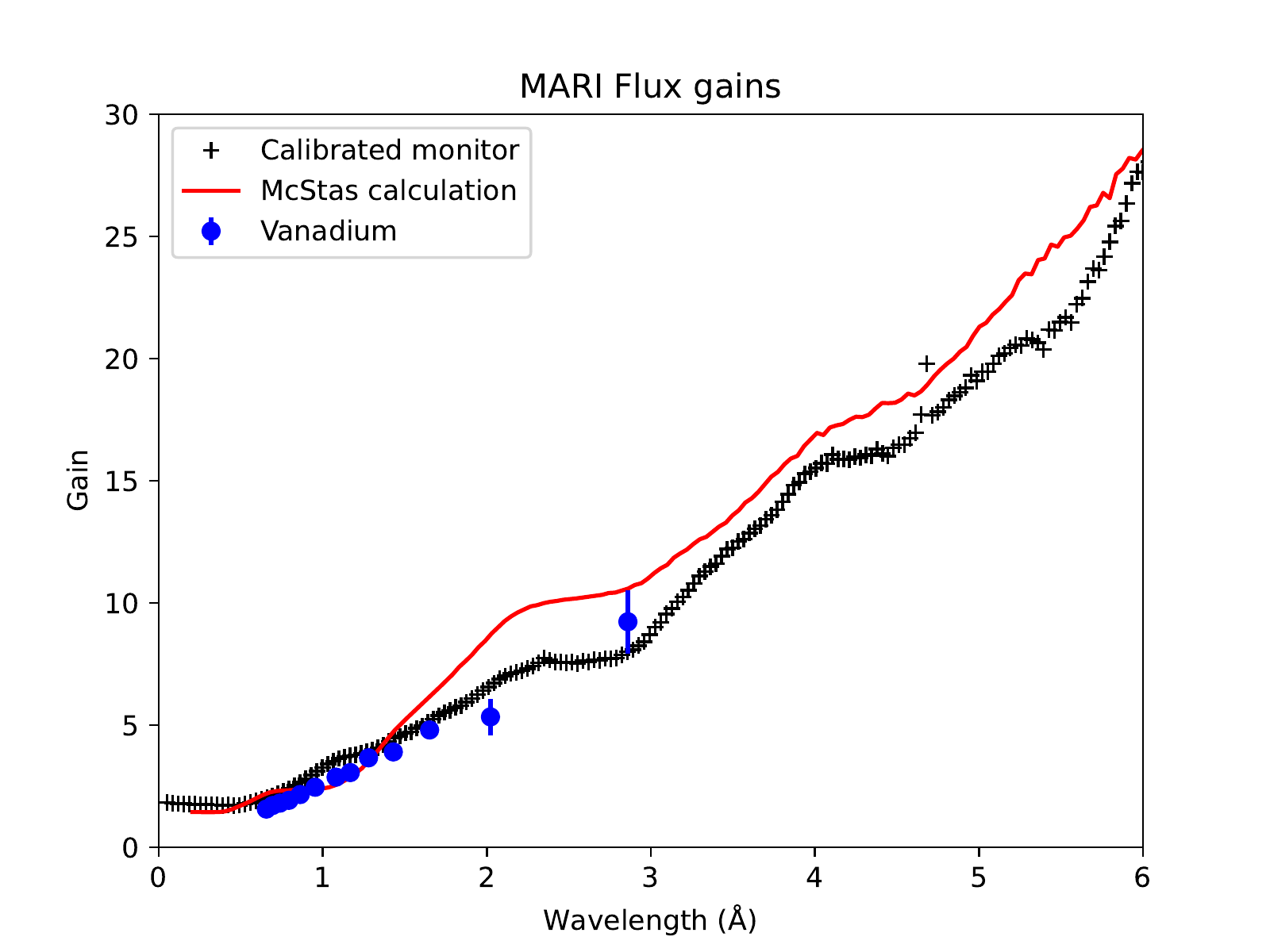}
    \caption{Gains from the new guide on MARI. Crosses show measurements with a calibrated monitor and
    circles are integrated scattered intensities from a Vanadium standard sample with the Gd chopper, averaged over runs with different frequencies at the same incident energy.
    The solid line is calculated from McStas simulations.}
    \label{fg:gains}
  \end{center}
\end{figure}

\begin{figure}[ht!]
  \begin{center}
    \includegraphics[width=\columnwidth, viewport=66 75 522 758]{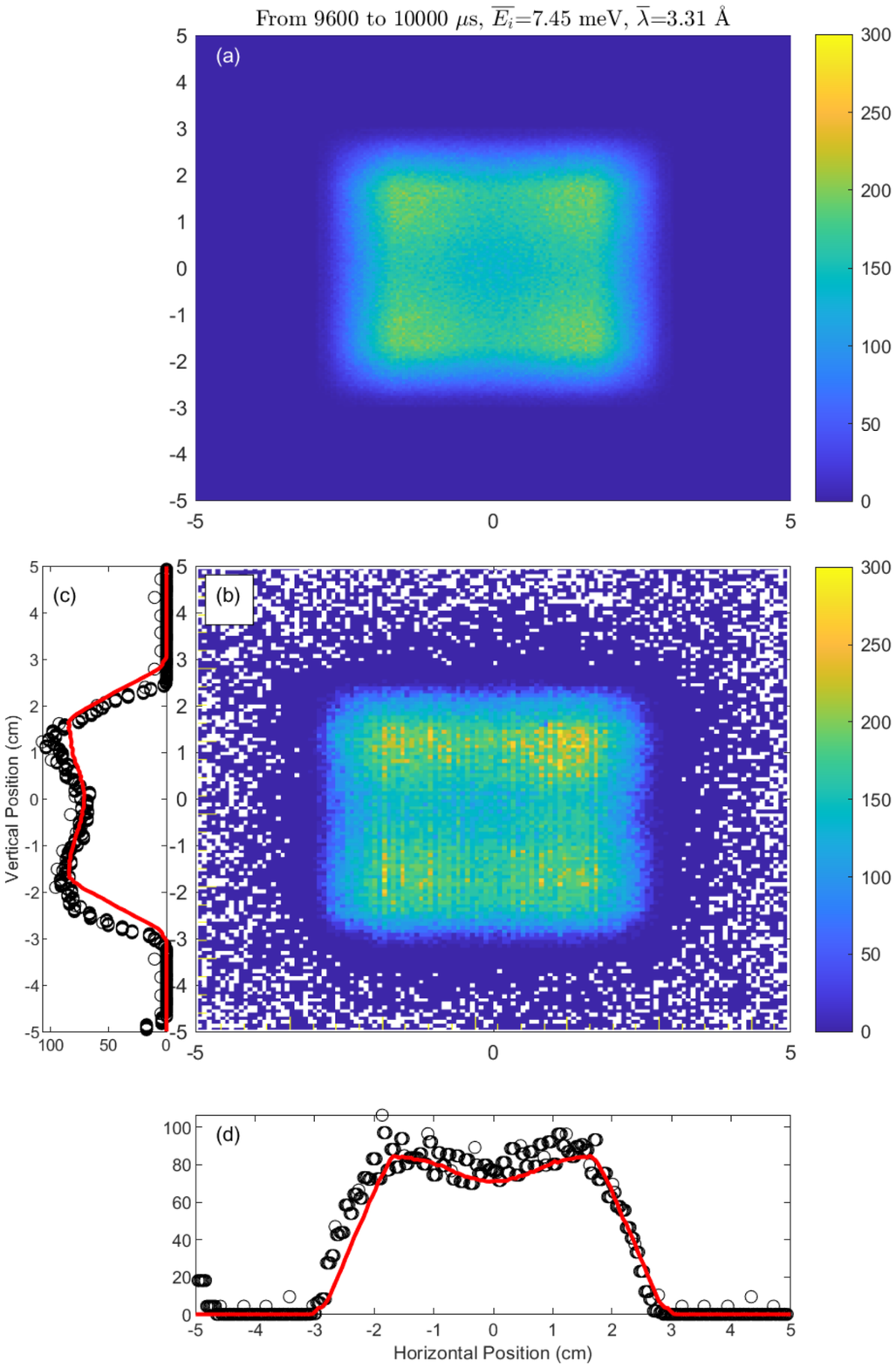}
    \caption{MARI beam-profile after the upgrade, as calculated using McStas (a) and measured with a 2D (nGEM) detector at the sample position (b), together with the vertical (c) and horizontal (d) beam profiles. The nGEM measurements and McStas calculated profiles are both in arbitrary units and have been scaled to match each other.}
    \label{fg:ngem}
  \end{center}
\end{figure}

\cref{fg:absflux} shows the measured and calculated absolute flux of the upgraded MARI instrument and~\cref{fg:gains} shows the measured and calculated gains from the guide.
The measurements, taken in 2016 before and in 2018 after the guide upgrade, used the same calibrated GS1 glass scintillating bead monitor~\cite{bewley_pooley} mounted at the sample position, giving an absolute flux measurement, and the gain was obtained by the ratio of the two results.
Likewise, the calculated gain was obtained from the ratio of the calculated flux from McStas~\cite{mcstas_jrs} simulations of the pre- and post-upgrade instrument.
The calculated flux in~\cref{fg:absflux} is shown scaled down by a factor of 2, which from our experience is necessary due to inaccuracies in the moderator models, based on MCNPX calculations~\cite{skoro2018}, used for ISIS instruments.

The measured and calculated flux gain is displayed in~\cref{fg:gains}, showing that around the peak flux ($\approx 1.8$~\AA) the flux gain from the upgrade is a factor of 6, and this increases with longer incident neutron wavelengths: At the lowest commonly used incident wavelength $\lambda$=3.7~\AA~(incident energy $E_i$=6~meV), the flux gain is a factor of 14. 
This has made low energy measurements more feasible than before the upgrade.
Test measurements were also made with $\lambda$=6.4~\AA~($E_i$=2~meV) where the gain is a factor of 30 but even so the flux is too low to be usable except with a large sample.

The flat plateaus in the gain are due to the gaps in the guide to accommodate the disk/Nimonic and Fermi chopper pits, which means that neutrons which would have been reflected by a guide there are lost.
These gaps also cause a slight non-uniformity in the beam profile for longer wavelengths, as shown in~\cref{fg:ngem}, where a dip in flux in the centre of the beam can be seen for a representative case of $\lambda=3.3$~\AA.
The full dataset of beam profiles for wavelengths from 1 to 10~\AA~is included in the supplementary materials.
The beam profile was measured using a 2D time-sensitive neutron gas electron multiplier (nGEM) detector~\cite{ngem} at the sample position with an active area of $100\times 100$mm and pixel size of 0.8~mm ($128 \times 128$ pixels).
Scaled McStas simulations integrated over the same time-of-flight ranges show good agreement with the measurements, as shown in~\cref{fg:ngem}.
The slightly non-uniform beam profile may result in a non-analytic resolution profile, and work is progressing to include a ray-tracing element to the resolution calculations used in standard analysis programs like Horace~\cite{ewings2016horace}.

\section{New disk chopper and RRM-mode} \label{sec-choppers}

\begin{figure}[ht!]
  \begin{center}
    \includegraphics[width=0.8\columnwidth, viewport=0 0 510 938]{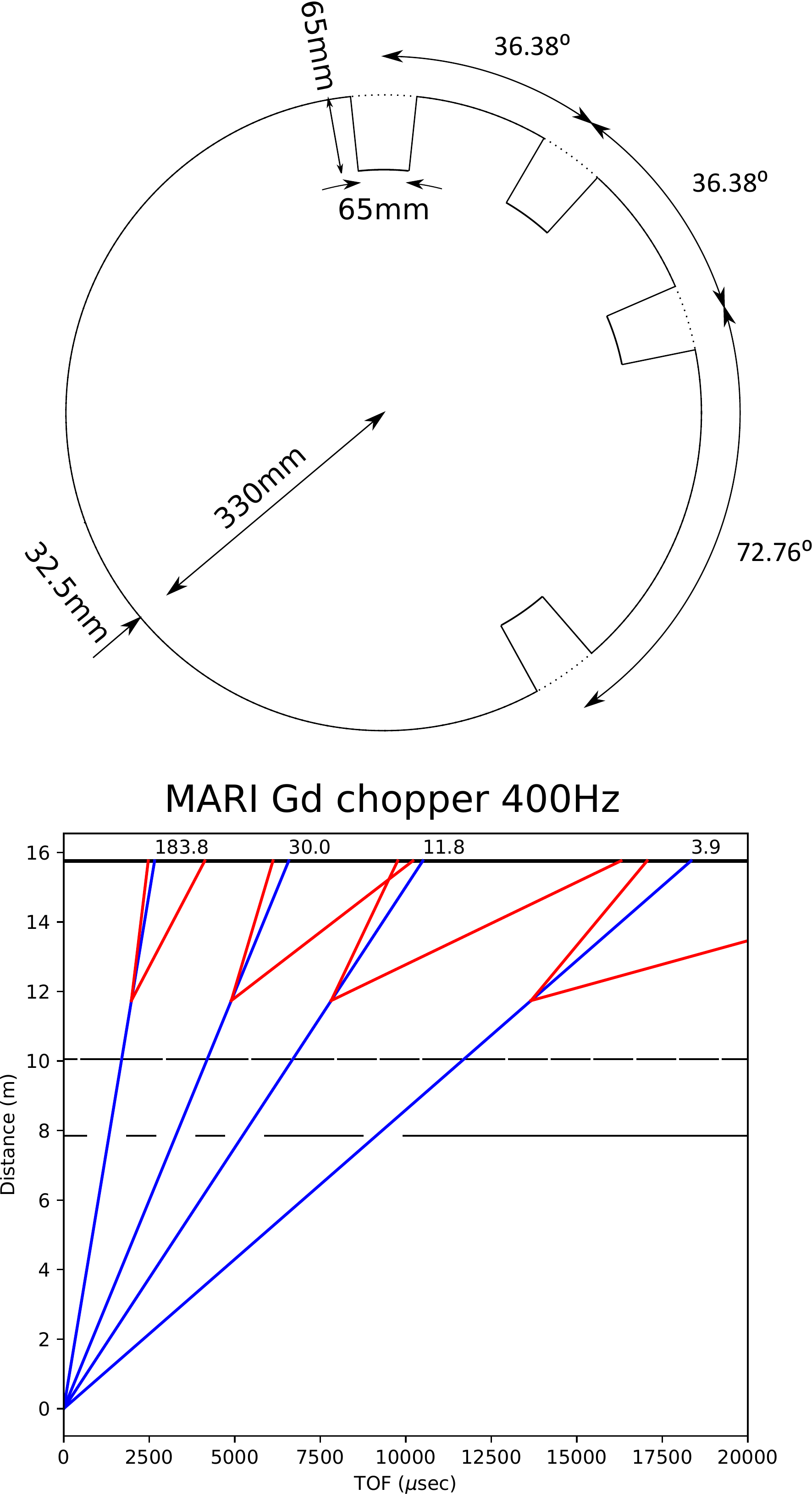}
    \caption{The MARI disk chopper design and time-distance diagram for a representative configuration. Red lines denote the time range where inelastic scattering from a neutron energy gain of 300~K (left boundary) to a neutron energy loss of 90\% of $E_i$ (right boundary) would fall.}
    \label{fg:disk}
  \end{center}
\end{figure}

In addition to the new guide, MARI has also had a new counter-rotating double-disk chopper installed, which allows the repetition rate multiplication (RRM) technique~\cite{russina_rrm} to be used to measure multiple incident energy pulses simultaneously. RRM mode on MARI is designed to be used with the Fermi chopper running at 200, 400 or 600~Hz. For example at 400~Hz, every 2.5~ms the Fermi chopper will make a complete (2$\pi$) rotation, and hence transmit a pulse of neutrons. Thus 8 neutron pulses will be transmitted within the 20~ms window between proton pulses (a time ``frame''). 
However, not all these pulses will be usable because the inelastic scattering from some of the pulses may overlap with other pulses. To prevent this ``pulse-overlap'', the disk chopper has 4 slots designed to pass pulses corresponding to zero (the principal pulse), one (``$2\pi$''), two (``$4\pi$'') or four (``$8\pi$'') full rotations of the Fermi chopper at 400~Hz, as shown in~\cref{fg:disk}. The curved boron chopper packages can only be used in RRM mode when run at 400~Hz, whilst the straight Gd chopper package is able to transmit a neutron pulse after half a rotation (a ``$\pi$'' pulse), so may be used at 200 or 600~Hz as well.
In addition, the two disk choppers may be phased to transmit a single or pairs of neutron pulses per proton pulse rather than all four, to allow more flexibility in operations. Transmitting a single pulse offers a lower background at the cost of being able to measure less, whilst transmitting two pulses (either the principal and ``$2\pi$'' or the principal and ``$4\pi$'' pulses at 400~Hz) may help with avoiding pulse-overlap.

\section{Detector electronics upgrade} \label{sec-det}

Concurrent with the guide upgrade, the detector electronics (but not the $^3$He detector tubes) were upgraded.
This consisted of new pre-amps, analogue-to-digital converters (ADC) and data-acquisition electronics (DAE).

Previously each tube on MARI was directly connected to its own pre-amp, housed in its own box, which was glued to the end of the tube. 
This meant that replacing faulty units took significant time and effort, and could only be done during a beam-off period.
To ease maintenance we decided to replace this with PCB-mounted pre-amps where the pre-amps for sets of eight tubes are housed in a single box.
Each PCB contains sixteen pre-amps so that a tube could be switched using a relay between one of two pre-amps in case of failure during a cycle.
In addition each pre-amp box also houses environmental sensors for ambient temperature and humidity to enable preemptive maintenance should high temperatures or humidity be detected.

\begin{figure}[t!]
  \begin{center}
    \includegraphics[width=0.918\columnwidth, viewport=123 279 468 556]{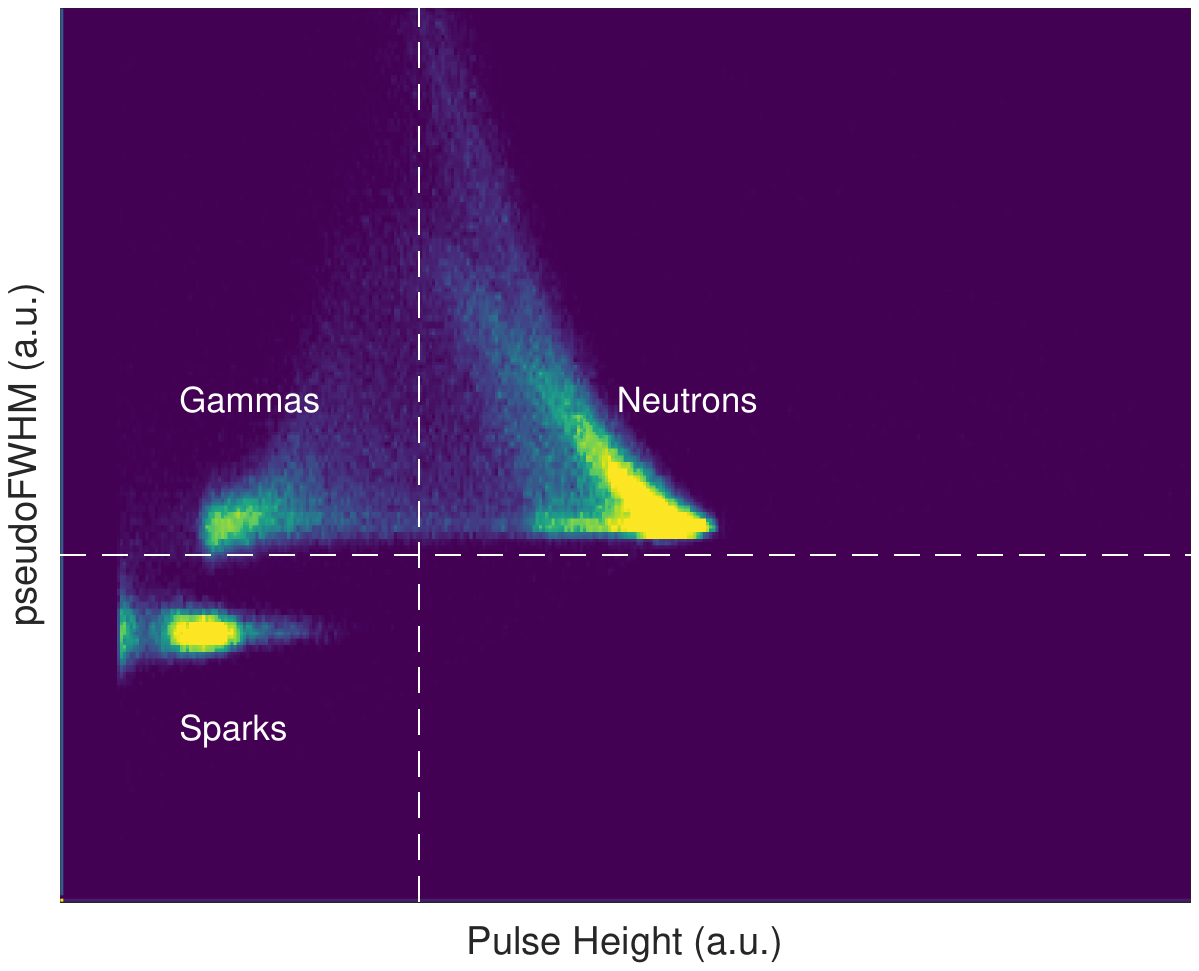}
    \caption{Histogram of pseudoFWHM vs pulse-height computed from traces collected using an Acquiris digitizer, showing how neutron signals can be distinguished from signals from gamma radiation and electrical noise (``sparks'').}
    \label{fg:spark-rejection}
  \end{center}
\end{figure}

The new design, however, requires a short length of wire to connect the tubes to the new pre-amp which was not needed in the old design.
During commissioning, it was discovered that the thin (3.1~mm diameter) standard copper wire used for this was prone to sparking which introduced large electrical background signals that can be mistaken for neutron counts by the DAE.
The thin wire was originally chosen because they would need minimal modification of the ends of the tube as this only permitted a relatively small diameter wire.
To solve this issue, we pursued two separate approaches: we investigated alternative wires for the link, and attempted to electronically distinguish the spark signal from neutron (or gamma) signals.
The second approach was only possible because the new ADC is built around a field-programmable gate array (FPGA) which allows fast digital signal processing to be carried out within the unit and easy and fast updating of this firmware by ``flashing'' the unit.
In the end, both methods for handling sparks were implemented: the thin wire was replaced by a much thicker (5.7~mm diameter) silicone-rubber sheathed high-voltage wire with the tube connectors re-designed and also a spark-rejection algorithm was incorporated into the ADC.

\begin{figure}[t!]
  \begin{center}
    \includegraphics[width=0.9\columnwidth, viewport=13 5 421 607]{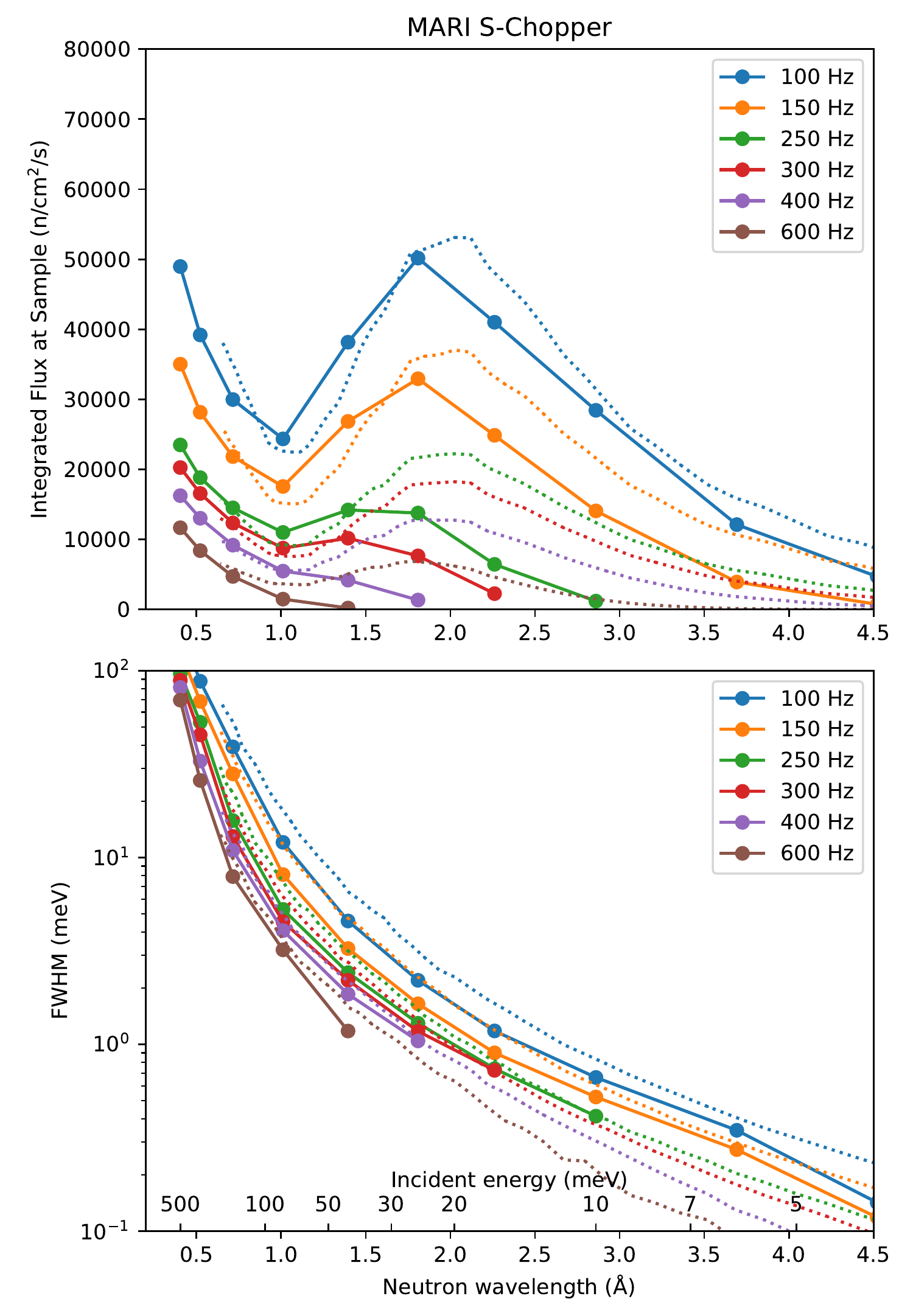}
    \caption{Measured (solid lines and circles) flux and linewidth (FWHM) at the sample and scaled calculated (dashed lines) flux and linewidth (FWHM) with the boron S chopper.}
    \label{fg:resflux_S}
  \end{center}
\end{figure}

\begin{figure}[ht!]
  \begin{center}
    \includegraphics[width=0.9\columnwidth, viewport=13 5 421 607]{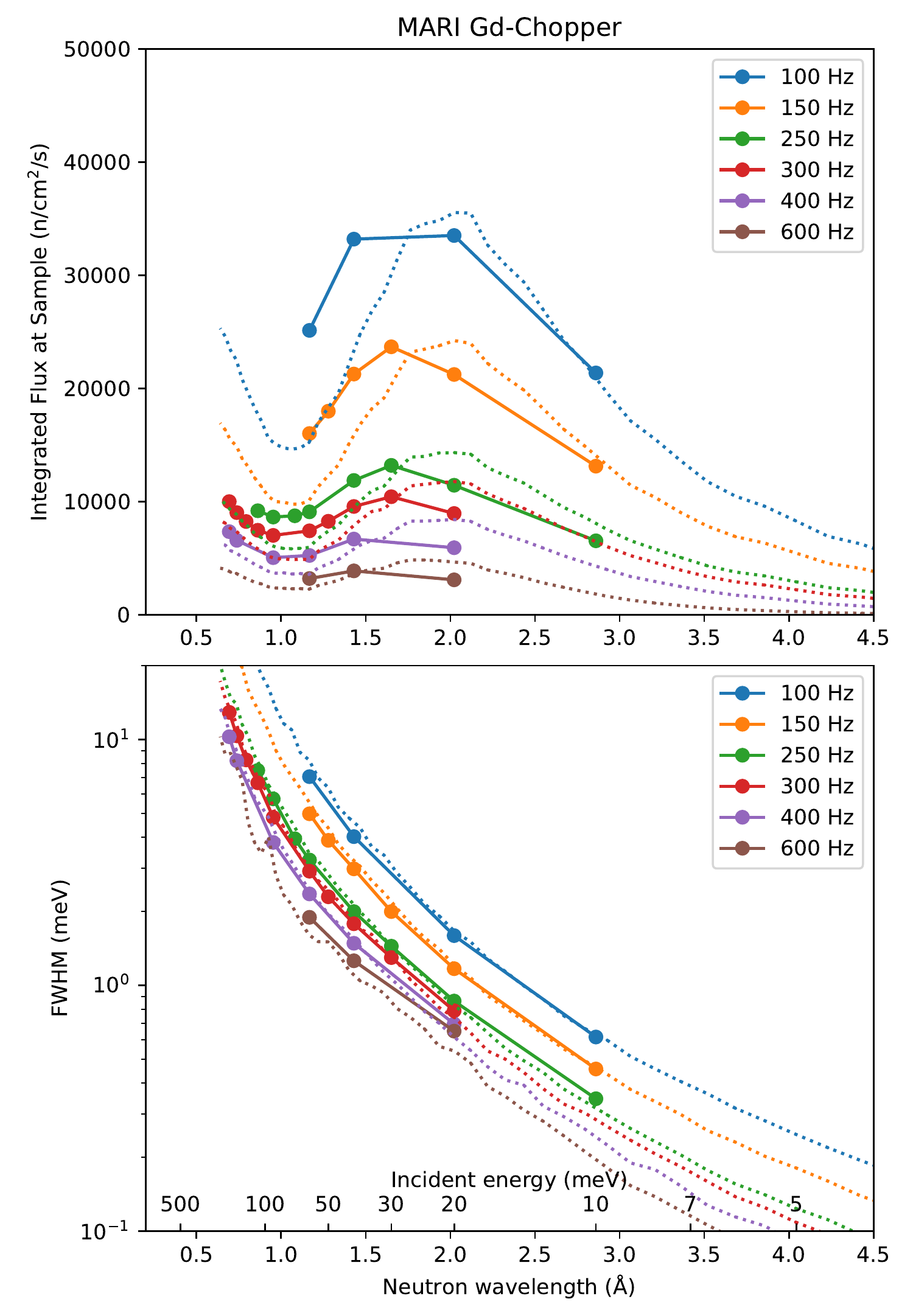}
    \caption{Measured (solid lines and circles) flux and linewidth (FWHM) at the sample and scaled calculated (dashed lines) flux and linewidth (FWHM) with the gadolinium chopper.}
    \label{fg:resflux_gd}
  \end{center}
\end{figure}

We now briefly discuss the spark-rejection algorithm, which relies on the spark signal being much more transient (shorter duration) than the response of the tube to ionisation by neutrons or gammas.
With the FPGA, we can compute a ``pseudoFWHM'' of the time response of the signal as:

\begin{equation}
    \mathrm{pseudoFWHM} = -V_0 \left/ \left( \left .
        \frac{\partial^2 V}{\partial t^2} \right|_{t=t_0} \right) \right.
\end{equation}

\noindent where $V_0$ is the peak amplitude of the pulse and $t_0$ is the time of the centre of the peak.
This quantity is actually proportional to the square of the full width at half maximum if the signal is a Gaussian.
We ensure that the signal is close to Gaussian by applying low-pass filters before computing the ``pseudoFWHM'', and avoid the costly square root computation which we found had little effect on the ability to discriminate sparks from other signals.
The second derivative at the peak centre is computed by finite difference of the digitised signal.

\cref{fg:spark-rejection} shows an example 2D histogram of the signal amplitude vs ``pseudoFWHM''.
The figure shows we can introduce two thresholds: a lower pulse-height threshold below which signals are considered to be gammas, and a lower pseudoFWHM threshold below which signals are considered to be sparks.

\section{Instrument performance} \label{sec-performance}

\begin{figure}[ht!]
  \begin{center}
    \includegraphics[width=\columnwidth, viewport=87 263 508 579]{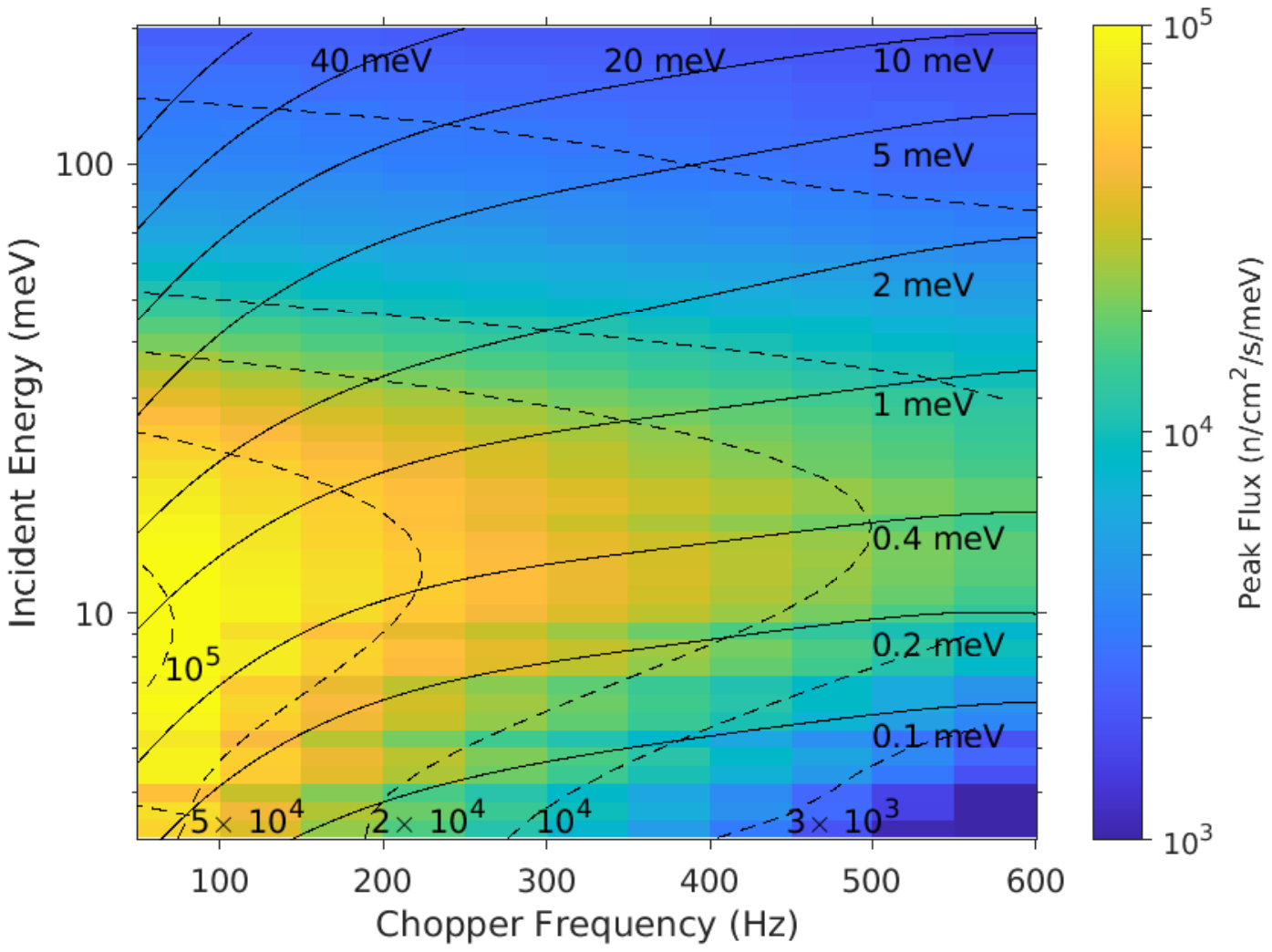}
    \caption{Energy resolution at the detector (solid contours) and peak flux at the sample position (colormap and dashed contours) with the Gd chopper on MARI, from McStas simulations.}
    \label{fg:resflux}
  \end{center}
\end{figure}

In order to verify the performance of the spectrometer after the upgrade, we performed extensive measurements with a standard annular vanadium sample. From this the integrated flux at a given incident energy and chopper frequency could be determined from the measured counts using the known sample mass and cross-sections. \cref{fg:resflux_S} and~\cref{fg:resflux_gd} show the measured and calculated integrated flux at the sample and vanadium elastic linewidths for the boron ``S'' (denoting medium or ``Sloppy'' resolution) and gadolinium choppers respectively.

~\cref{fg:resflux} shows the (calculated) expected performance of MARI with the gadolinium chopper in an ``at-a-glance" format, including both the \emph{calculated} energy resolution and \emph{calculated} peak flux as a function of incident energy and chopper frequency simultaneously, with contours indicating the resolution and colours the flux.

\section{Science Highlights} \label{sec-science}

Since the instrument was recommissioned in 2018, over 90 experiments have been carried out on MARI (including short ``Xpress'' experiments to characterise new samples). About two-thirds of these experiments are on magnetic materials, of which half are measurements of ordered spin waves, one third are crystalline electric field measurements and the remainder are on disordered systems, or magnetic clusters.

\begin{figure}[ht!]
  \begin{center}
    \includegraphics[width=\columnwidth, viewport=3 5 357 278]{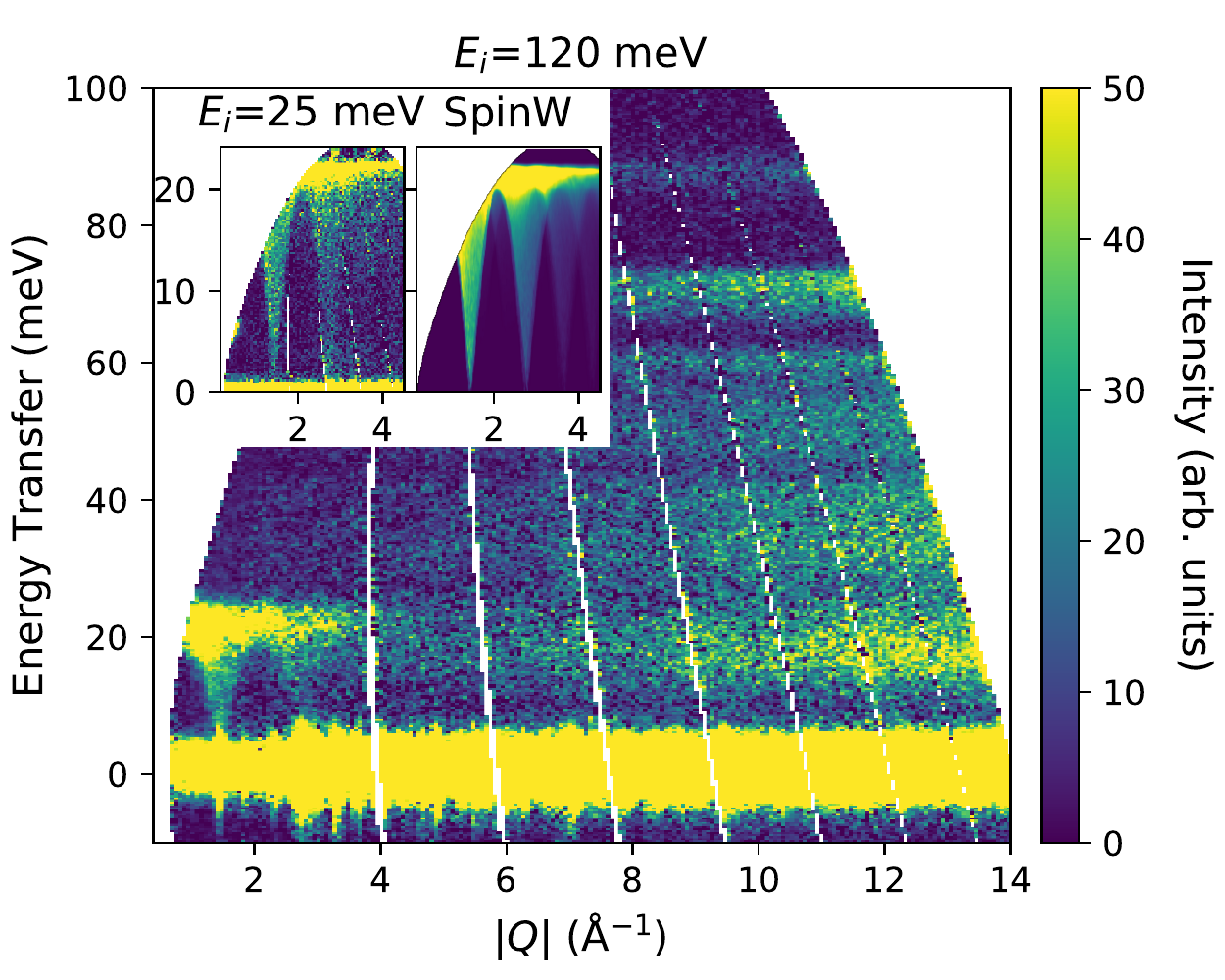}
    \caption{Inelastic neutron spectra of YCrO$_3$ at 5~K with the empty sample container background subtracted. The inset shows the second ``rep'' at a lower incident energy which reveals details of the spin wave excitations confronted by a model linear spin wave calculation as described by Bajaj et al~\cite{Bajaj_2021}.}
    \label{fg:ycro3}
  \end{center}
\end{figure}

MARI's wide angular coverage ($5^{\circ} < 2\theta <= 135^{\circ}$) and ability to use repetition rate multiplication to measure several incident energies together is useful for simultaneous measurements of phonons and magnons, as exemplified by a recent work on the multiferroic material YCrO$_3$~\cite{Bajaj_2021}.
\cref{fg:ycro3} shows a single 3 hour measurement with the Gd chopper running at 400~Hz that captures data at both $E_i$=120~meV and $E_i$=25~meV.
The $E_i$=120~meV data clearly shows the phonon modes at high $|Q|$, whilst the $E_i$=25~meV pulse shows details of the spin wave spectrum which can now be readily modelled with programs such as \texttt{SpinW}~\cite{toth_2015}.

\begin{figure}[ht!]
  \begin{center}
    \includegraphics[width=\columnwidth, viewport=16 38 484 475]{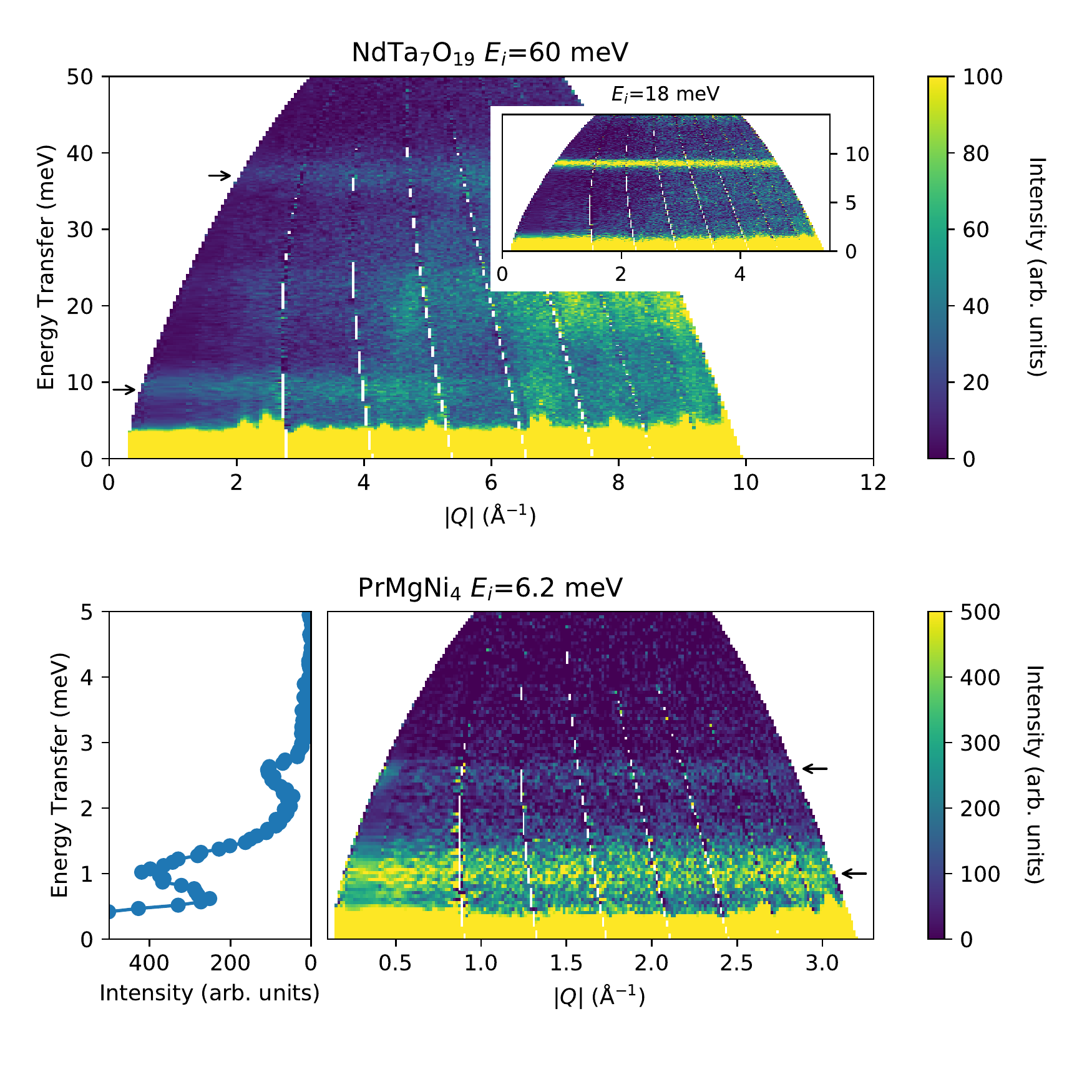}
    \caption{Inelastic neutron spectra of (top) NdTa$_7$O$_{19}$ at 5~K with $E_i$=60~meV and (top inset) $E_i$=18~meV, and (bottom) PrMgNi$_4$ showing crystalline electric field transitions as indicated by the arrows. 
    The scattering around 20~meV at high $|Q|$ in the NdTa$_7$O$_{19}$ $E_i$=60~meV data is from Aluminium phonons from the sample container.
    (Bottom left) A line cut integrated over the range $0 < |Q| < 2$~\AA$^{-1}$ showing crystalline electric field excitations at 1 and 2.6~meV in PrMgNi$_4$.}
    \label{fg:cef}
  \end{center}
\end{figure}

While they are a decreasing fraction of experiments on MARI, crystalline electric field (CEF) measurements still account for a significant amount of beam time, with measurements on traditional strongly correlated intermetallics such as PrMgNi$_4$~\cite{kusanose_2022} and on newer geometrically frustrated oxides such as the quantum spin-liquid candidate NdTa$_7$O$_{19}$~\cite{arh_2022}, both shown in~\cref{fg:cef}.

\begin{figure}[ht!]
  \begin{center}
    \includegraphics[width=\columnwidth, viewport=7 0 351 283]{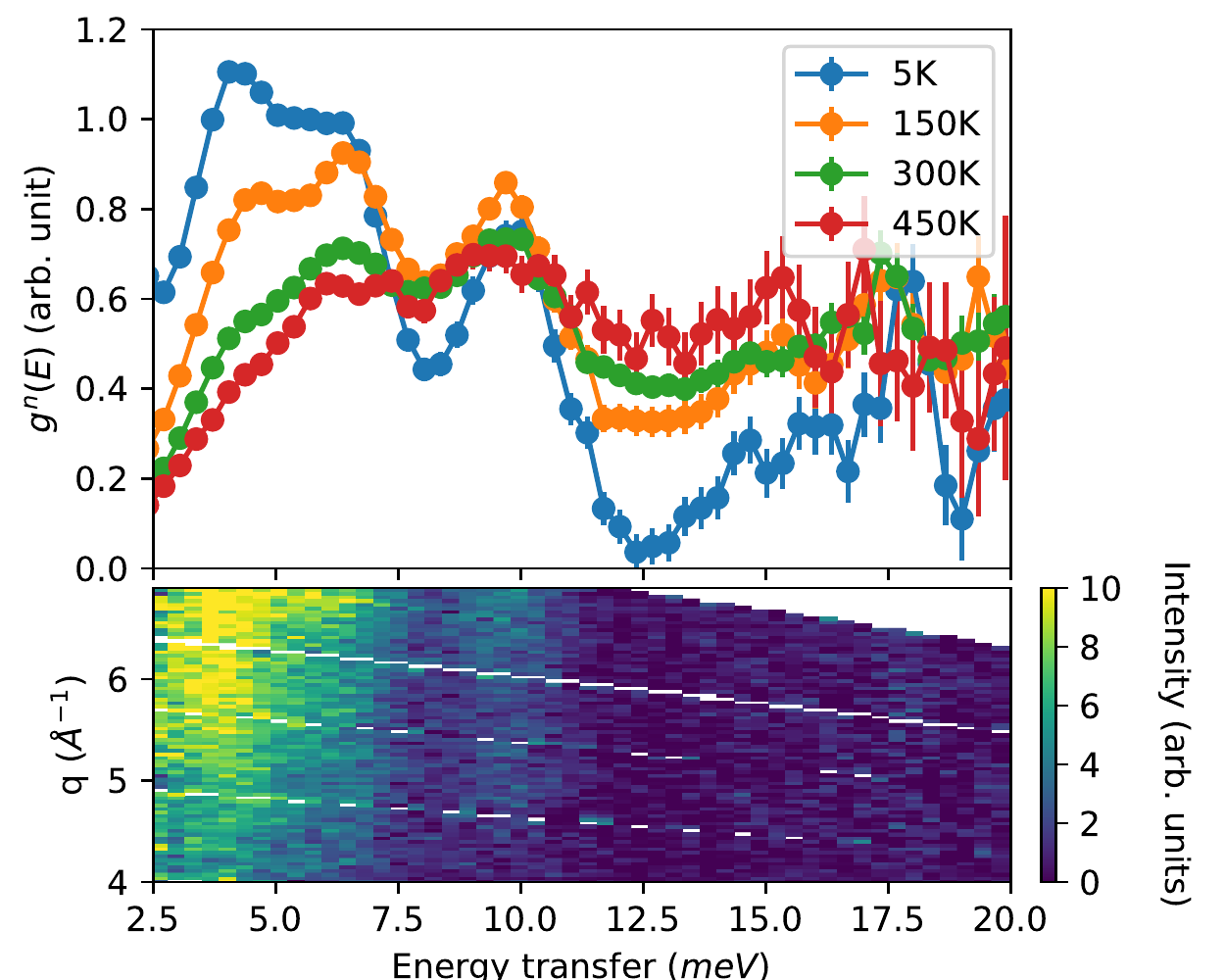}
    \caption{Neutron weighted phonon density of state $g^n(E)$ as a function of temperature (upper panel) and powder inelastic neutron spectra (lower panel) of TlInTe$_2$ at 5~K.}
    \label{fg:tlinte2}
  \end{center}
\end{figure}

In the case of PrMgNi$_4$, the cubic point group is expected to split the $J=4$ ground multiplet into two triplets, a doublet and a singlet, with the $\Gamma_3$ doublet being the ground state, and yielding two dipole allowed transitions at the lowest temperatures.
Instead, the experiment observed three transitions, at 11.6 (not shown), 2.5 and 1.1~meV.
These two lower energy transitions indicate a splitting of the $\Gamma_3$ doublet, likely due to a symmetry lowering caused by Pr-Mg inter-site disorder, which in turn prevents a quadrupolar ordering seen in similar cubic praseodymium intermetallics~\cite{kusanose_2022}.
Prior to the upgrade, MARI would have struggled to measure the 1.1~meV level as we would have needed to use higher incident energies with worse resolution. 
The increased flux at lower energies from the guide upgrade instead allowed us to use a lower incident energy and to thus clearly resolve the mode.

NdTa$_7$O$_{19}$ is a triangular lattice antiferromagnet and quantum spin liquid candidate where no magnetic order has been observed above 66~mK. Crucially it shows no evidence of the site disorder that complicates other spin liquid candidates. Neutron scattering measurements and fits of the Nd crystal fields reveal a strong Ising anisotropy in agreement with ESR data, but with some residual $xy$ character. The dominant Ising anisotropy together with the triangular geometry is thought to prevent magnetic ordering, whilst the residual $xy$ interactions acts as a perturbing transverse field which could lead to a zero temperature phase transition~\cite{arh_2022}.

Finally, although magnetic studies have been a staple of experiments on MARI historically, a rising proportion of experiments are now looking at phonons in functional materials, such as batteries cathode materials, ionic conductors or thermoelectrics. One recent example is the Zintl phase TlInTe$_2$~\cite{dutta_2021}, whose INS spectrum and neutron-weighted phonon density of states (PDOS) is shown in~\cref{fg:tlinte2}. This potential thermoelectric material has an ultra-low thermal conductivity, $\approx$0.4 W m$^{-1}$ K$^{-1}$ above room temperature. Measurements on MARI together with an X-ray pair distribution function study showed that this is due to highly anharmonic rattling of Tl ions inside InTe$_2$ cages which causes strong phonon scattering and hence strongly suppresses the thermal conductivity. The rattling is anisotropic and leads to several low energy modes which can be seen at around 3.5, 5.6 and 9.5~meV in~\cref{fg:tlinte2} which broaden dramatically as the temperature rises confirming their anharmonicity.

\section{Summary} \label{sec-summary}

The MARI spectrometer at ISIS has been upgraded with new $m=3$ supermirror guides giving a factor of 6 gain in flux at $\lambda=1.8$~\AA ($E=25$~meV) where the peak flux from the methane moderator is, whilst larger gains at lower energies enable new measurements which were previously unfeasible.

The detector electronics were also upgraded with a new FPGA-based ADC to allow fast and flexible digital signal processing.

Finally, the upgrade also includes a new double disk chopper which allows routine use of repetition rate multiplication without rep-overlap to improve experimental throughput.
For example the full dispersion bandwidths of phonon or magnon excitations can be measured by one incident energy neutron pulse whilst details of the dispersion or possible energy gaps can be probed simultaneously with another incident energy pulse.

The upgrades have together allowed MARI to continue to deliver a high quality science program well into its fourth decade, covering a wide range of topics from quantum magnetism to functional materials.

\section*{Acknowledgments}

The guide and detector electronics upgrade of MARI was made possible thanks to the efforts of numerous ISIS staff involved in various aspects of the project. We are grateful for their effort and support that enabled the project to be completed successfully.

\bibliographystyle{apsrev4-1}
\bibliography{refs} 

\begin{thebibliography}{12}%
\makeatletter
\providecommand \@ifxundefined [1]{%
 \@ifx{#1\undefined}
}%
\providecommand \@ifnum [1]{%
 \ifnum #1\expandafter \@firstoftwo
 \else \expandafter \@secondoftwo
 \fi
}%
\providecommand \@ifx [1]{%
 \ifx #1\expandafter \@firstoftwo
 \else \expandafter \@secondoftwo
 \fi
}%
\providecommand \natexlab [1]{#1}%
\providecommand \enquote  [1]{``#1''}%
\providecommand \bibnamefont  [1]{#1}%
\providecommand \bibfnamefont [1]{#1}%
\providecommand \citenamefont [1]{#1}%
\providecommand \href@noop [0]{\@secondoftwo}%
\providecommand \href [0]{\begingroup \@sanitize@url \@href}%
\providecommand \@href[1]{\@@startlink{#1}\@@href}%
\providecommand \@@href[1]{\endgroup#1\@@endlink}%
\providecommand \@sanitize@url [0]{\catcode `\\12\catcode `\$12\catcode
  `\&12\catcode `\#12\catcode `\^12\catcode `\_12\catcode `\%12\relax}%
\providecommand \@@startlink[1]{}%
\providecommand \@@endlink[0]{}%
\providecommand \url  [0]{\begingroup\@sanitize@url \@url }%
\providecommand \@url [1]{\endgroup\@href {#1}{\urlprefix }}%
\providecommand \urlprefix  [0]{URL }%
\providecommand \Eprint [0]{\href }%
\providecommand \doibase [0]{http://dx.doi.org/}%
\providecommand \selectlanguage [0]{\@gobble}%
\providecommand \bibinfo  [0]{\@secondoftwo}%
\providecommand \bibfield  [0]{\@secondoftwo}%
\providecommand \translation [1]{[#1]}%
\providecommand \BibitemOpen [0]{}%
\providecommand \bibitemStop [0]{}%
\providecommand \bibitemNoStop [0]{.\EOS\space}%
\providecommand \EOS [0]{\spacefactor3000\relax}%
\providecommand \BibitemShut  [1]{\csname bibitem#1\endcsname}%
\let\auto@bib@innerbib\@empty
\bibitem [{\citenamefont {Taylor}\ \emph {et~al.}(1990)\citenamefont {Taylor},
  \citenamefont {Arai}, \citenamefont {Bennington}, \citenamefont {Bowden},
  \citenamefont {Osborn}, \citenamefont {Andersen}, \citenamefont {Stirling},
  \citenamefont {Nakane}, \citenamefont {Yamada},\ and\ \citenamefont
  {Welz}}]{taylor90}%
  \BibitemOpen
  \bibfield  {author} {\bibinfo {author} {\bibfnamefont {A.~D.}\ \bibnamefont
  {Taylor}}, \bibinfo {author} {\bibfnamefont {M.}~\bibnamefont {Arai}},
  \bibinfo {author} {\bibfnamefont {S.~M.}\ \bibnamefont {Bennington}},
  \bibinfo {author} {\bibfnamefont {Z.~A.}\ \bibnamefont {Bowden}}, \bibinfo
  {author} {\bibfnamefont {R.}~\bibnamefont {Osborn}}, \bibinfo {author}
  {\bibfnamefont {K.}~\bibnamefont {Andersen}}, \bibinfo {author}
  {\bibfnamefont {W.~G.}\ \bibnamefont {Stirling}}, \bibinfo {author}
  {\bibfnamefont {T.}~\bibnamefont {Nakane}}, \bibinfo {author} {\bibfnamefont
  {K.}~\bibnamefont {Yamada}}, \ and\ \bibinfo {author} {\bibfnamefont
  {D.}~\bibnamefont {Welz}},\ }\href
  {http://www.neutronresearch.com/parch/1990/01/199001007050.pdf} {\bibfield
  {journal} {\bibinfo  {journal} {Proc. ICANS-XI, KEK, Tsukuba}\ ,\ \bibinfo
  {pages} {705}} (\bibinfo {year} {1990})}\BibitemShut {NoStop}%
\bibitem [{\citenamefont {Bewley}\ and\ \citenamefont
  {Pooley}(2022)}]{bewley_pooley}%
  \BibitemOpen
  \bibfield  {author} {\bibinfo {author} {\bibfnamefont {R.}~\bibnamefont
  {Bewley}}\ and\ \bibinfo {author} {\bibfnamefont {D.}~\bibnamefont
  {Pooley}},\ }\href {\doibase https://doi.org/10.1016/j.nima.2022.167161}
  {\bibfield  {journal} {\bibinfo  {journal} {Nucl. Instrum. Methods Phys. Res.
  A}\ }\textbf {\bibinfo {volume} {1039}},\ \bibinfo {pages} {167161} (\bibinfo
  {year} {2022})}\BibitemShut {NoStop}%
\bibitem [{\citenamefont {Willendrup}\ and\ \citenamefont
  {Lefmann}(2020)}]{mcstas_jrs}%
  \BibitemOpen
  \bibfield  {author} {\bibinfo {author} {\bibfnamefont {P.~K.}\ \bibnamefont
  {Willendrup}}\ and\ \bibinfo {author} {\bibfnamefont {K.}~\bibnamefont
  {Lefmann}},\ }\href {\doibase 10.3233/JNR-190108} {\bibfield  {journal}
  {\bibinfo  {journal} {Journal of Neutron Research}\ }\textbf {\bibinfo
  {volume} {22}},\ \bibinfo {pages} {1} (\bibinfo {year} {2020})}\BibitemShut
  {NoStop}%
\bibitem [{\citenamefont {Škoro}\ \emph {et~al.}(2018)\citenamefont {Škoro},
  \citenamefont {Lilley},\ and\ \citenamefont {Bewley}}]{skoro2018}%
  \BibitemOpen
  \bibfield  {author} {\bibinfo {author} {\bibfnamefont {G.}~\bibnamefont
  {Škoro}}, \bibinfo {author} {\bibfnamefont {S.}~\bibnamefont {Lilley}}, \
  and\ \bibinfo {author} {\bibfnamefont {R.}~\bibnamefont {Bewley}},\ }\href
  {\doibase https://doi.org/10.1016/j.physb.2017.12.060} {\bibfield  {journal}
  {\bibinfo  {journal} {Physica B: Condensed Matter}\ }\textbf {\bibinfo
  {volume} {551}},\ \bibinfo {pages} {381} (\bibinfo {year} {2018})},\ \bibinfo
  {note} {the 11th International Conference on Neutron Scattering (ICNS
  2017)}\BibitemShut {NoStop}%
\bibitem [{\citenamefont {Ohshita}\ \emph {et~al.}()\citenamefont {Ohshita},
  \citenamefont {Ishiwata}, \citenamefont {Iwase}, \citenamefont {Fujisaki},
  \citenamefont {Muto}, \citenamefont {Satoh}, \citenamefont {Seya},
  \citenamefont {Sakaguchi}, \citenamefont {Otomo}, \citenamefont {Ikeda},
  \citenamefont {Kaneko},\ and\ \citenamefont {Suzuya}}]{ngem}%
  \BibitemOpen
  \bibfield  {author} {\bibinfo {author} {\bibfnamefont {H.}~\bibnamefont
  {Ohshita}}, \bibinfo {author} {\bibfnamefont {M.}~\bibnamefont {Ishiwata}},
  \bibinfo {author} {\bibfnamefont {K.}~\bibnamefont {Iwase}}, \bibinfo
  {author} {\bibfnamefont {F.}~\bibnamefont {Fujisaki}}, \bibinfo {author}
  {\bibfnamefont {S.}~\bibnamefont {Muto}}, \bibinfo {author} {\bibfnamefont
  {S.}~\bibnamefont {Satoh}}, \bibinfo {author} {\bibfnamefont
  {T.}~\bibnamefont {Seya}}, \bibinfo {author} {\bibfnamefont {M.}~\bibnamefont
  {Sakaguchi}}, \bibinfo {author} {\bibfnamefont {T.}~\bibnamefont {Otomo}},
  \bibinfo {author} {\bibfnamefont {K.}~\bibnamefont {Ikeda}}, \bibinfo
  {author} {\bibfnamefont {N.}~\bibnamefont {Kaneko}}, \ and\ \bibinfo {author}
  {\bibfnamefont {K.}~\bibnamefont {Suzuya}},\ }\enquote {\bibinfo {title} {New
  neutron beam monitor based on {GEM}},}\ in\ \href {\doibase
  10.7566/JPSCP.8.036019} {\emph {\bibinfo {booktitle} {Proceedings of the 2nd
  International Symposium on Science at J-PARC — Unlocking the Mysteries of
  Life, Matter and the Universe —}}}\BibitemShut {NoStop}%
\bibitem [{\citenamefont {Ewings}\ \emph {et~al.}(2016)\citenamefont {Ewings},
  \citenamefont {Buts}, \citenamefont {Le}, \citenamefont {Van~Duijn},
  \citenamefont {Bustinduy},\ and\ \citenamefont {Perring}}]{ewings2016horace}%
  \BibitemOpen
  \bibfield  {author} {\bibinfo {author} {\bibfnamefont {R.}~\bibnamefont
  {Ewings}}, \bibinfo {author} {\bibfnamefont {A.}~\bibnamefont {Buts}},
  \bibinfo {author} {\bibfnamefont {M.}~\bibnamefont {Le}}, \bibinfo {author}
  {\bibfnamefont {J.}~\bibnamefont {Van~Duijn}}, \bibinfo {author}
  {\bibfnamefont {I.}~\bibnamefont {Bustinduy}}, \ and\ \bibinfo {author}
  {\bibfnamefont {T.}~\bibnamefont {Perring}},\ }\href {\doibase
  10.1016/j.nima.2016.07.036} {\bibfield  {journal} {\bibinfo  {journal}
  {Nuclear Instruments and Methods in Physics Research Section A: Accelerators,
  Spectrometers, Detectors and Associated Equipment}\ }\textbf {\bibinfo
  {volume} {834}},\ \bibinfo {pages} {132} (\bibinfo {year}
  {2016})}\BibitemShut {NoStop}%
\bibitem [{\citenamefont {Russina}\ \emph {et~al.}(2012)\citenamefont
  {Russina}, \citenamefont {Mezei},\ and\ \citenamefont {Kali}}]{russina_rrm}%
  \BibitemOpen
  \bibfield  {author} {\bibinfo {author} {\bibfnamefont {M.}~\bibnamefont
  {Russina}}, \bibinfo {author} {\bibfnamefont {F.}~\bibnamefont {Mezei}}, \
  and\ \bibinfo {author} {\bibfnamefont {G.}~\bibnamefont {Kali}},\ }\href
  {\doibase 10.1088/1742-6596/340/1/012018} {\bibfield  {journal} {\bibinfo
  {journal} {Journal of Physics: Conference Series}\ }\textbf {\bibinfo
  {volume} {340}},\ \bibinfo {pages} {012018} (\bibinfo {year}
  {2012})}\BibitemShut {NoStop}%
\bibitem [{\citenamefont {Bajaj}\ \emph {et~al.}(2021)\citenamefont {Bajaj},
  \citenamefont {Roy}, \citenamefont {Khandelwal}, \citenamefont
  {Chattopadhyay}, \citenamefont {Sathe}, \citenamefont {Mishra}, \citenamefont
  {Mittal}, \citenamefont {Babu}, \citenamefont {Le}, \citenamefont
  {Niedziela},\ and\ \citenamefont {Bansal}}]{Bajaj_2021}%
  \BibitemOpen
  \bibfield  {author} {\bibinfo {author} {\bibfnamefont {N.}~\bibnamefont
  {Bajaj}}, \bibinfo {author} {\bibfnamefont {A.~P.}\ \bibnamefont {Roy}},
  \bibinfo {author} {\bibfnamefont {A.}~\bibnamefont {Khandelwal}}, \bibinfo
  {author} {\bibfnamefont {M.~K.}\ \bibnamefont {Chattopadhyay}}, \bibinfo
  {author} {\bibfnamefont {V.}~\bibnamefont {Sathe}}, \bibinfo {author}
  {\bibfnamefont {S.~K.}\ \bibnamefont {Mishra}}, \bibinfo {author}
  {\bibfnamefont {R.}~\bibnamefont {Mittal}}, \bibinfo {author} {\bibfnamefont
  {P.~D.}\ \bibnamefont {Babu}}, \bibinfo {author} {\bibfnamefont {M.~D.}\
  \bibnamefont {Le}}, \bibinfo {author} {\bibfnamefont {J.~L.}\ \bibnamefont
  {Niedziela}}, \ and\ \bibinfo {author} {\bibfnamefont {D.}~\bibnamefont
  {Bansal}},\ }\href {\doibase 10.1088/1361-648x/abd781} {\bibfield  {journal}
  {\bibinfo  {journal} {Journal of Physics: Condensed Matter}\ }\textbf
  {\bibinfo {volume} {33}},\ \bibinfo {pages} {125702} (\bibinfo {year}
  {2021})}\BibitemShut {NoStop}%
\bibitem [{\citenamefont {Toth}\ and\ \citenamefont {Lake}(2015)}]{toth_2015}%
  \BibitemOpen
  \bibfield  {author} {\bibinfo {author} {\bibfnamefont {S.}~\bibnamefont
  {Toth}}\ and\ \bibinfo {author} {\bibfnamefont {B.}~\bibnamefont {Lake}},\
  }\href {\doibase 10.1088/0953-8984/27/16/166002} {\bibfield  {journal}
  {\bibinfo  {journal} {Journal of Physics: Condensed Matter}\ }\textbf
  {\bibinfo {volume} {27}},\ \bibinfo {pages} {166002} (\bibinfo {year}
  {2015})}\BibitemShut {NoStop}%
\bibitem [{\citenamefont {Kusanose}\ \emph {et~al.}(2022)\citenamefont
  {Kusanose}, \citenamefont {Onimaru}, \citenamefont {Yamane}, \citenamefont
  {Umeo}, \citenamefont {Takabatake}, \citenamefont {Guidi}, \citenamefont
  {Le},\ and\ \citenamefont {Adroja}}]{kusanose_2022}%
  \BibitemOpen
  \bibfield  {author} {\bibinfo {author} {\bibfnamefont {Y.}~\bibnamefont
  {Kusanose}}, \bibinfo {author} {\bibfnamefont {T.}~\bibnamefont {Onimaru}},
  \bibinfo {author} {\bibfnamefont {Y.}~\bibnamefont {Yamane}}, \bibinfo
  {author} {\bibfnamefont {K.}~\bibnamefont {Umeo}}, \bibinfo {author}
  {\bibfnamefont {T.}~\bibnamefont {Takabatake}}, \bibinfo {author}
  {\bibfnamefont {T.}~\bibnamefont {Guidi}}, \bibinfo {author} {\bibfnamefont
  {D.}~\bibnamefont {Le}}, \ and\ \bibinfo {author} {\bibfnamefont
  {D.}~\bibnamefont {Adroja}},\ }\href {\doibase
  10.1088/1742-6596/2164/1/012052} {\bibfield  {journal} {\bibinfo  {journal}
  {Journal of Physics: Conference Series}\ }\textbf {\bibinfo {volume}
  {2164}},\ \bibinfo {pages} {012052} (\bibinfo {year} {2022})}\BibitemShut
  {NoStop}%
\bibitem [{\citenamefont {Arh}\ \emph {et~al.}(2022)\citenamefont {Arh},
  \citenamefont {Sana}, \citenamefont {Pregelj}, \citenamefont {Khuntia},
  \citenamefont {Jagličić}, \citenamefont {Le}, \citenamefont {Biswas},
  \citenamefont {Manuel}, \citenamefont {Mangin-Thro}, \citenamefont
  {Ozarowski},\ and\ \citenamefont {Zorko}}]{arh_2022}%
  \BibitemOpen
  \bibfield  {author} {\bibinfo {author} {\bibfnamefont {T.}~\bibnamefont
  {Arh}}, \bibinfo {author} {\bibfnamefont {B.}~\bibnamefont {Sana}}, \bibinfo
  {author} {\bibfnamefont {M.}~\bibnamefont {Pregelj}}, \bibinfo {author}
  {\bibfnamefont {P.}~\bibnamefont {Khuntia}}, \bibinfo {author} {\bibfnamefont
  {Z.}~\bibnamefont {Jagličić}}, \bibinfo {author} {\bibfnamefont {M.~D.}\
  \bibnamefont {Le}}, \bibinfo {author} {\bibfnamefont {P.~K.}\ \bibnamefont
  {Biswas}}, \bibinfo {author} {\bibfnamefont {P.}~\bibnamefont {Manuel}},
  \bibinfo {author} {\bibfnamefont {L.}~\bibnamefont {Mangin-Thro}}, \bibinfo
  {author} {\bibfnamefont {A.}~\bibnamefont {Ozarowski}}, \ and\ \bibinfo
  {author} {\bibfnamefont {A.}~\bibnamefont {Zorko}},\ }\href {\doibase
  10.1038/s41563-021-01169-y} {\bibfield  {journal} {\bibinfo  {journal}
  {Nature Materials}\ }\textbf {\bibinfo {volume} {21}},\ \bibinfo {pages}
  {416} (\bibinfo {year} {2022})}\BibitemShut {NoStop}%
\bibitem [{\citenamefont {Dutta}\ \emph {et~al.}(2021)\citenamefont {Dutta},
  \citenamefont {Samanta}, \citenamefont {Ghosh}, \citenamefont {Voneshen},\
  and\ \citenamefont {Biswas}}]{dutta_2021}%
  \BibitemOpen
  \bibfield  {author} {\bibinfo {author} {\bibfnamefont {M.}~\bibnamefont
  {Dutta}}, \bibinfo {author} {\bibfnamefont {M.}~\bibnamefont {Samanta}},
  \bibinfo {author} {\bibfnamefont {T.}~\bibnamefont {Ghosh}}, \bibinfo
  {author} {\bibfnamefont {D.~J.}\ \bibnamefont {Voneshen}}, \ and\ \bibinfo
  {author} {\bibfnamefont {K.}~\bibnamefont {Biswas}},\ }\href {\doibase
  https://doi.org/10.1002/anie.202013923} {\bibfield  {journal} {\bibinfo
  {journal} {Angewandte Chemie International Edition}\ }\textbf {\bibinfo
  {volume} {60}},\ \bibinfo {pages} {4259} (\bibinfo {year}
  {2021})}\BibitemShut {NoStop}%
\end{thebibliography}%

\end{document}